\documentclass[prd,showpacs,showkeys]{revtex4}
\usepackage{amsmath,amsthm,amsfonts,amssymb}
\usepackage[mathcal]{eucal}
\usepackage{mathrsfs}
\usepackage{graphicx}
\usepackage{dcolumn}
\usepackage{latexsym}
\usepackage{epsfig}
\usepackage{bm}
\usepackage[all]{xy}
\newcommand{\ie}{\begin{equation}}
\newcommand{\fe}{\end{equation}}

\newcommand\fverb{\setbox\fverbbox=\hbox\bgroup\verb}
\newcommand\fverbdo{\egroup\medskip\noindent%
            \fbox{\unhbox\fverbbox}\ }
\newcommand\fverbit{\egroup\item[\fbox{\unhbox\fverbbox}]}
\newbox\fverbbox

\def\text#1{\mbox{#1}}

\begin{document}
\title{Gauge vector field localization on 3-brane placed in a warped transverse resolved conifold}
\author{F. W. V. Costa$\;^{a,b}$}
\author{J. E. G. Silva$\;^{a}$}
\author{C. A. S. Almeida$\;^{a}$}
\address{$\;^{a}$Departamento de F\'{\i}sica, Universidade Federal do
Cear\'a \\ Caixa Postal 6030, CEP 60455-760, Fortaleza, Cear\'a, Brazil}
\address{$\;^{b}$FAFIDAM, Universidade Estadual do Cear\'a \\ Limoeiro do Norte, Cear\'a, Brazil}

\begin{abstract}
We have investigated the features of the gauge vector field in a braneworld scenario built as a warped product between a 3-brane and a $2$-cycle of the resolved conifold. This scenario allowed us to study how the gauge field behaves when the transverse manifold evolves upon a geometric flow that controls the singularity at the origin. Besides, since the transverse manifold has a cylindrical symmetry according to the $3$-brane, this geometry can be regarded as a near brane correction of the string-like branes. Indeed, by means of a new warp function and the angular metric component of the resolved conifold, the braneworld can exhibit a conical form near the origin as well as a regular behavior in that region. The analysis of the gauge field in this background has been carried out for the $s-$wave state and a normalizable massless mode was found. For the massive modes, the resolution parameter avoids an infinite well on the brane and controls the depth of the well and the high of the barrier around the brane. The
massive modes are slightly changed near the brane but they agreed with the string-like spectrum for large distances.

\end{abstract}

\maketitle


\section{Introduction}

The Kaluza-Klein theories, as well as other extra dimension theories, has been boosted after the seminal papers of Randall and Sundrum (RS) \cite{Randall:1999ee,Randall:1999vf}.
By assuming an specific geometry, namely, a warped product between a flat 3-brane and a single extra dimension, the RS models solved the hierarchy problem, providing a tiny correction to the
Newtonian potential and allowing gravity propagation in an extra dimension.

Indeed, RS models opened new trends to theoretical research. Soon after, a huge amount of papers appeared enhancing features of RS models as proving the stability of the geometrical solution \cite{Goldberger:1999uk}, providing a physical source for this geometry \cite{Csaki:2000fc,Gremm:1999pj,Bazeia:2004dh,Almeida:2009jc}, allowing the other fields to propagate in the bulk \cite{Goldberger:1999wh,Kehagias:2000au,Liu:2007gk,Huber:2000ie} or extending the model to higher dimensions \cite{Chodos:1999zt,Liu:2007gk,Olasagasti:2000gx,Chen:2000at,Cohen:1999ia,Gregory:1999gv,Gherghetta:2000qi,Oda:2000zc,Giovannini:2001hh,Ponton:2000gi,Kehagias:2004fb,Benson:2001ac,Oda2000a}.

In six dimensions, some models assume  two extra dimensions (an extension of the so called RS type 1 model) where some two dimensional compact transverse manifold has been proposed \cite{Chodos:1999zt}, including some peculiar manifolds as the torus \cite{Duan:2006es}, an apple shape
space \cite{Gogberashvili:2007gg} and a football shaped manifold \cite{Garriga:2004tq}. On the other hand, some authors have studied a braneworld where the two dimensional manifold has a cylindrical symmetry according to the 3-brane, the so-called string-like brane \cite{Chen:2000at,Cohen:1999ia,Gherghetta:2000qi,Giovannini:2001hh,Oda:2000zc,Ponton:2000gi,Gregory:1999gv,Liu:2007gk,Olasagasti:2000gx,Kehagias:2004fb,Tinyakov:2001jt}.
This brane can be regarded as a $4$-dimensional vortex embedded in a $6$-dimensional space-time \cite{Cohen:1999ia,Giovannini:2001hh,Gregory:1999gv,Olasagasti:2000gx}.

Among the good achievements of the last models are the localization of the fermion \cite{Liu:2007gk} and gauge \cite{Oda:2000zc} fields on the brane coupled only with gravity and a lower correction to the gravitational potential \cite{Gherghetta:2000qi}. Besides, the geometry is
richer than the $5-$D RS type 2 model, since the exterior space-time of the string-brane is conical with deficit angle proportional to the string tensions \cite{Gherghetta:2000qi,Giovannini:2001hh,Chen:2000at,Olasagasti:2000gx}. Hence,
the exterior geometry of the string-brane reflects the physical content of the brane.

This relationship between geometry and physics led us to study how the physical characteristics of a braneworld change when the geometry of the transverse manifold evolves under a geometrical flow. This flux can be realized as a symmetry change (perhaps a break) of the geometry due to variations of the parameters that describe the brane.

In order to accomplish this task we have chosen as a transverse manifold a 2-cycle of a well-known space in string theory, the so-called resolved conifold. This is a smooth parameter-dependent six
dimensional space whose parameter $a$ controls the singularity on the tip of the cone \cite{Candelas:1989js,p,Minasian:1999tt,Greene:1995hu,Cvetic:2000mh,VazquezPoritz:2001zt,Klebanov:2007us}. Thus, it is
possible continuously flow from a smooth to a singular manifold by means of variations of the parameter $a$.

The resolved conifold is one of the smoothed manifolds from the conifold, a Calabi-Yau (CY) orbifold that plays an important role in the conical transitions in string theory
\cite{Candelas:1989js,p,Greene:1995hu,Minasian:1999tt,Papadopoulos:2000gj,Klebanov:1999tb,Klebanov:2000hb,Klebanov:2000nc,Klebanov:2007us,Pando Zayas:2000sq}. There is another smoothed conifold, obtained by deforming through a parameter the quadric that defines the conifold, called deformed conifold \cite{Cvetic:2000mh,Candelas:1989js,Firouzjahi:2005qs,Greene:1995hu,Klebanov:2000hb,Noguchi:2005ws,Pando Zayas:2000sq}.
Both spaces are quite important in some extensions of the AdS-CFT correspondence, where the branes are placed at the nobe of the smoothed cones \cite{Klebanov:1999tb,Klebanov:2000hb,Klebanov:2000nc,Klebanov:2007us,Pando Zayas:2000sq}.

Moreover, the behavior of the gravitational field on a bulk builded from a warped product between a flat $3$-brane and a deformed or resolved conifold has already been studied. For the former case, the authors studied the gravity in the whole deformed conifold or in a compact domain nearby a region of the deformed conifold called Klebanov-Strassler throat \cite{Brummer:2005sh,Firouzjahi:2005qs,Noguchi:2005ws}. It has been argued that the massless mode, usually identified with the effective gravity, is normalizable and the Kaluza-Klein (KK) modes are peaked in the throat. For the resolved conifold, despite there is an infinite radial dimension, the massless mode is located around the origin and the KK spectrum has an exponential decay \cite{VazquezPoritz:2001zt}.

The study of the behavior of the fields on braneworlds with non trivial transverse manifolds has already been addressed in the literature. Indeed, for a manifold like a tear drop, the conical singularity plays an important role in adjusting the brane cosmological constant \cite{Kehagias:2004fb}. Further, Ricci-flat or homogeneous spaces locate not only gravity but also chiral fermions \cite{RandjbarDaemi:2000ft}. Another singular solution is the so-called cigar-like universe where the transverse space has a cylindrical shape far from the brane but the radius of the cylinder shrinks as we move toward the brane \cite{deCarlos:2003nq}.

In this article we are concerned with an axisymmetric and static six dimensional braneworld with cosmological constant whose transverse space is a $2$-cycle of the resolved conifold. We have investigated
its geometrical issues, as well the behavior of the vector gauge field in this background. In our last article \cite{Silva:2011yk}, we have shown that a real scalar field exhibits some interesting results in this geometry, as a parametrization of the well-known volcano potential for the KK modes and a robust localization of the massless mode upon the resolution flow. Here, through a new warp function that possess a $Z_{2}$ symmetry, we have analyzed the components of stress-energy-momentum tensor, the string tensions and the relation between the mass scales. Although near the origin the scalar curvature depends strongly upon the resolution parameter, the bulk converges to a $AdS_{6}$ manifold, regardless the value of the resolution parameter $a$.

We argue that this is a realistic scenario since the components of the energy-momentum tensor satisfy the weak energy condition and the failure of the dominant energy condition is shared with other six dimensional models, as the Gherghetta-Schaposhnikov (GS) model \cite{Gherghetta:2000qi}, where the $3$-brane is infinitely thin \cite{Tinyakov:2001jt}. Nevertheless, the energy string tension (string mass) is always greater than the other string tension, as in GS model \cite{Gherghetta:2000qi}. Besides, when the value of the resolution parameter increases the relationship between the mass scales increases also.

For $a=0$, this geometry can be realized as a complete (interior and exterior) string-like solutions that provides some corrections to the thin string model near the origin. On the other hand, for $a\neq 0$, the 3-brane can be regarded as a brane embedded in a 4-brane with a compact extra dimension whose radius is the resolution parameter. This enable us to realize the RS type 1 model
as a limit of the six dimensional non-compact scenario.

In addition, we have studied how the gauge field behaves under the resolution flow. Due to the non-trivial geometry, we assumed that the brane component of the gauge field does not depend on the radial coordinate.  This condition yields a homogeneous differential equation for the KK modes which is valid only for the $s-$wave solution, i.e., for $l=0$. For this solution, the massless mode exhibits a $Z_{2}$ symmetry and it is trapped to the brane for any value of $a\neq 0$. For $a=0$, the massless mode vanishes on the brane due to the conical behavior. Therefore, the resolution parameter also allows to define the gauge zero-mode
on the brane, despite the conical characteristic. The problem of find the induced field equation on the brane also occurs with the gravitational field in
string-like modes \cite{Bostock:2003cv,Kanno:2004nr}.

Furthermore, the KK modes are well-behaved near the brane and they converge asymptotically to the well-known string-like spectrum \cite{Oda:2000zc}.
Besides, $a$ prevents an infinite well on the brane, which happens for conical spaces ($a=0$), and controls the high of the barrier and the depth of the well
around the brane.

This work is organized as follows. In section \ref{Conifold} we have review the most important features of the resolved conifold. Furthermore, we have proposed and studied the properties of the $2$-cycle of the resolved conifold that we have chosen as the transverse manifold. In section \ref{Bulk geometry} we built the warped product between a $3$-brane and the $2$-cycle described before. We presented and discussed the properties of the warp function chosen and the angular metric component. From the Einstein equation, we have studied the properties of the energy-momentum tensor components and the respective string tensions. Further, we analyzed the behavior of the relation between the mass scales with the resolution parameter.
In section \ref{Gauge vector field localization} we have obtained the massless mode as well as the KK spectrum for the $s-$wave solution. In this section we have studied also the response of the gauge field to the resolution flow. Some conclusions and perspectives are outlined in section \ref{Conclusions and perspectives}.


\section{Conifold geometry}
\label{Conifold}

In this section we present the definitions and main properties of the conifold as well of its smooth version, the so-called resolved manifold. Further, we choose and study
some characteristics of a 2-cycle of the resolved conifold that we shall use as a transverse manifold.

The 6-Conifold is a conical manifold $\mathcal{C}_{6}\subset \mathbb{C}^{4}$ defined as the solution of the quadric equation \cite{Candelas:1989js,p,Klebanov:2000hb,Pando Zayas:2000sq,Cvetic:2000mh,Firouzjahi:2005qs,Greene:1995hu,Klebanov:1999tb,Klebanov:2007us,Papadopoulos:2000gj,VazquezPoritz:2001zt}
\begin{equation}
\label{conifoldequation}
 z_{1}^{2}+z_{2}^{2}+z_{3}^{2}+z^{2}_{4}=0.
\end{equation}

A key property of this equation, inherent to the conics, is that, if $(z_{1},z_{2},z_{3},z_{4})$ satisfy eq. $(\ref{conifoldequation})$ then
$(\lambda z_{1},\lambda z_{2},\lambda z_{3},\lambda z_{4})$, $\lambda \in \mathbb{C}$, also satify eq. $(\ref{conifoldequation})$. The point $\lambda=0$ is the so-called node or tip of conifold whereas
for a fixed $\lambda$ we obtain a manifold called the base space $(X^{5})$.

By construction, it is possible to define a radial coordinate $u:[0,\infty) \rightarrow [0,\infty)$. Using such variable, a rather general metric of a 6-conifold over a $X^{5}$ compact space
takes the form
\begin{equation}
\label{conicalmetric}
 ds^{2}_{6}=du^{2}+u^{2}ds^{2}(X^{5}).
\end{equation}

For a well-known coset base space $X^{5}=T^{1,1}=SU(2)\times SU(2)/U(1)$, we can employ the coordinate system where $\theta_{1},\theta_{2} \in [0,\pi]$,
$\phi_{1},\phi_{2} \in [0,2\pi]$ and $\psi\in[0,4\pi]$ leading the metric of eq. (\ref{conicalmetric}) to \cite{Candelas:1989js,p,Klebanov:2000hb,Pando Zayas:2000sq}
\begin{eqnarray}
\label{conifoldmetric}
 ds^{2}_{6} & = & du^{2}+\frac{u^{2}}{9}(d\psi+\cos\theta_{1}d\phi_{1}+\cos\theta_{2}d\phi_{2})^{2}\nonumber\\
            & + & \frac{u^{2}}{6}(d\theta_{1}^{2}+\sin^{2}\theta_{1} d\phi_{1}^{2}+d\theta_{2}^{2}+\sin^{2}\theta_{2} d\phi_{2}^{2}).
\end{eqnarray}
This space has a naked singularity in $r=0$. A smooth version of this conifold, called the resolved conifold, is a parameter-dependent family of manifolds whose metrics can be written as
\cite{Candelas:1989js,Pando Zayas:2000sq,Cvetic:2000mh,VazquezPoritz:2001zt}
\begin{eqnarray}
\label{metricresolvedconifold}
 ds^{2}_{6} & = & \left(\frac{u^{2}+6a^{2}}{u^{2}+9a^{2}}\right)du^{2}+\frac{u^{2}}{9}\left(\frac{u^{2}+9a^{2}}{u^{2}+6a^{2}}\right)(d\psi+\cos\theta_{1}d\phi_{1}+\cos\theta_{2}d\phi_{2})^{2}\nonumber\\
            & + & \frac{1}{6}u^{2}(d\theta_{1}^{2}+\sin^{2}\theta_{1} d\phi_{1}^{2})+\frac{1}{6}(u^{2}+6a^{2})(d\theta_{2}^{2}+\sin^{2}\theta_{2} d\phi_{2}^{2}),
\end{eqnarray}
where, $a\in \mathbb{R}$ is a parameter with dimension $[a]=L$.
\begin{figure}
  \centering
\includegraphics[scale=1]{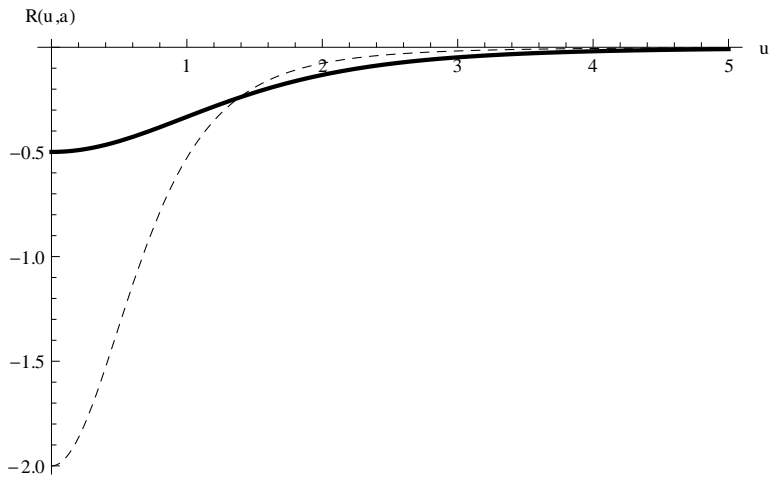}
 \caption{The scalar curvature of the 2-cycle of the resolved conifold for some values of $a$. The curvature is everywhere non-positive and it vanishes asymptotically. For $a=0.5$ (dashed
 line) the curvature is greater (in modulo) than for $a=1$ (thick line).}
\label{resolvedscalarcurvature}
\end{figure}
Since for $a=0$ we regain the singular conifold again, the parameter $a$ measure how smooth is the conifold and then it is called resolution parameter.

Note that in the limit $u\rightarrow 0$ the metric (\ref{metricresolvedconifold}) converges to a spherical one of radius $a$
\begin{equation}
 \lim_{u\rightarrow 0}{ds^{2}_{6}}=a^{2}(d\theta_{2}^{2}+\sin^{2}\theta_{2} d\phi_{2}^{2}),
\end{equation}
that has no singularity. Topologically this can be seen as a result of to take out a small neighborhood around $r=0$
and replaced it by a $S^{2}$ of radius $a$.

Since the metric for the coordinates $\psi,\phi_{1},\theta_{2},\phi_{2}$ vanishes near the origin, where the singularity is, let us concern ourselves with 2-cycle cone
that remains finite in the origin, namely
\begin{eqnarray}
\label{transversespacemetric}
 ds^{2}_{2} & = & \left(\frac{u^{2}+6a^{2}}{u^{2}+9a^{2}}\right)du^{2}+ \frac{1}{6}(u^{2}+6a^{2})d\theta^{2}.
\end{eqnarray}

This cone has a radial metric component $g_{uu}=\alpha(u,a)=\left(\frac{u^{2}+6a^{2}}{u^{2}+9a^{2}}\right)$.
Note that $\lim_{u\rightarrow \infty}{g_{uu}}=1$ and therefore the cone approaches asymptotically to the plane $\mathbb{R}^{2}$
with cylindrical metric of an effective radius $u_{eff}=\sqrt{\frac{(r^{2}+6a^{2})}{6}}$ which is the transverse metric
used in string-like geometries \cite{Cohen:1999ia,Gregory:1999gv,Gherghetta:2000qi,Giovannini:2001hh}. The angular resolved conifold metric component
$g_{\theta\theta}=\beta(u,a)=\frac{(u^{2}+6a^{2})}{6}$ has a conical singularity depending on
the resolution parameter.

The scalar curvature of this 2-manifold is
\begin{equation}
  R=R(r,a)=-\frac{6a^{2}(r^{2}+18a^{2})}{(r^{2}+6a^{2})^{3}}.
\end{equation}
As shown in the Fig. $(\ref{resolvedscalarcurvature})$, this $2$-section is everywhere smooth for $a\neq 0$. Since for $a=0$ this manifold is cone-like, its curvature vanishes except at the origin, where it diverges.

These issues motivated us to use this manifold as a prototype of extension of transverse
spaces in the brane worlds. As a matter of fact, many authors have studied the localization of fields in spherical backgrounds
whose transverse space has positive, constant and non-singular curvature \cite{Cohen:1999ia,Gregory:1999gv,Gherghetta:2000qi,Oda:2000zc,Olasagasti:2000gx,Giovannini:2001hh,Chodos:1999zt,Tinyakov:2001jt}. Other authors investigated the behavior of fields on other less common geometries \cite{deCarlos:2003nq,Firouzjahi:2005qs,Noguchi:2005ws,RandjbarDaemi:2000ft,Duan:2006es,Garriga:2004tq,Huber:2000ie}.
Since the resolved conifold is parameterized by the resolution parameter, by using resolved conifold as transverse manifold we can not only study the properties of the fields in this rich space as also we can study the effects that singular and smooth manifolds have on the localization of fields.

Now let us make a change of variables in order to write out the metric (\ref{transversespacemetric}) of $\mathcal{C}_{2}$ in a gaussian form
\begin{equation}
 ds^{2}_{2} = dr^2 + \beta(r,a)d\theta^{2}.
\end{equation}

By setting $dr=\sqrt{\frac{(u^2 + 6a^2)}{u^2 + 9a^2}}du$, we obtain a smooth change of variable $r: [0,\infty) \rightarrow [0,\infty)$ given by
\begin{equation}
\label{changeofvariable}
 r_{a}=\int^{u}{\sqrt{\frac{(u'^2 + 6a^2)}{u'^2 + 9a^2}}du'}.
\end{equation}

\begin{figure}[htb] 
       \begin{minipage}[b]{0.48 \linewidth}
           \fbox{\includegraphics[width=\linewidth]{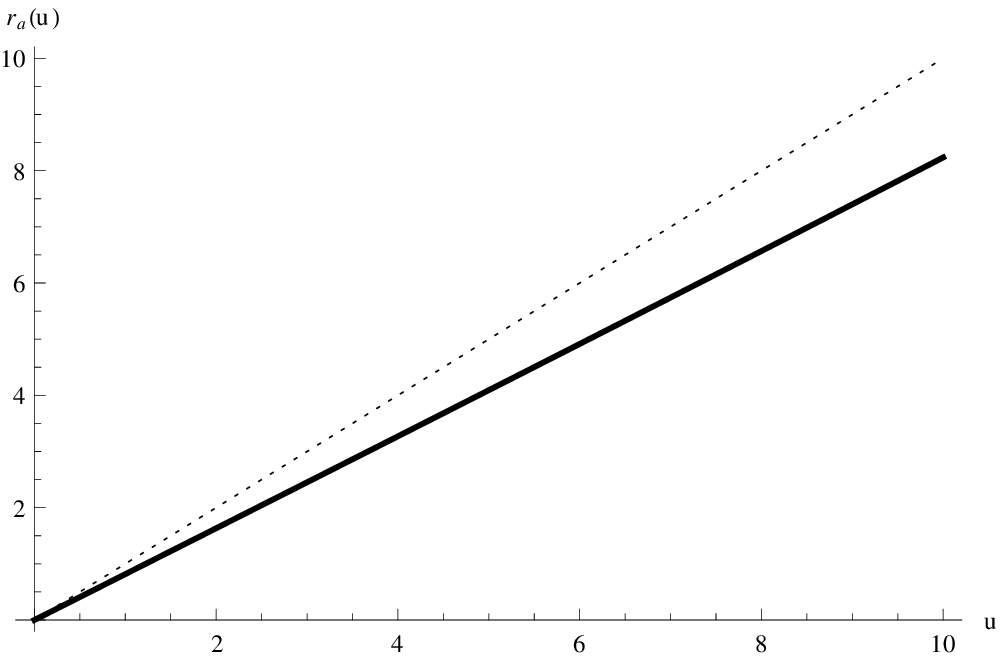}}\\
           \caption{Change of radial coordinate. The slope of the graphic for $a=10$ is less than for $a=0$ (dotted line).}
           \label{variablechange}
       \end{minipage}\hfill
       \begin{minipage}[b]{0.48 \linewidth}
           \fbox{\includegraphics[width=\linewidth]{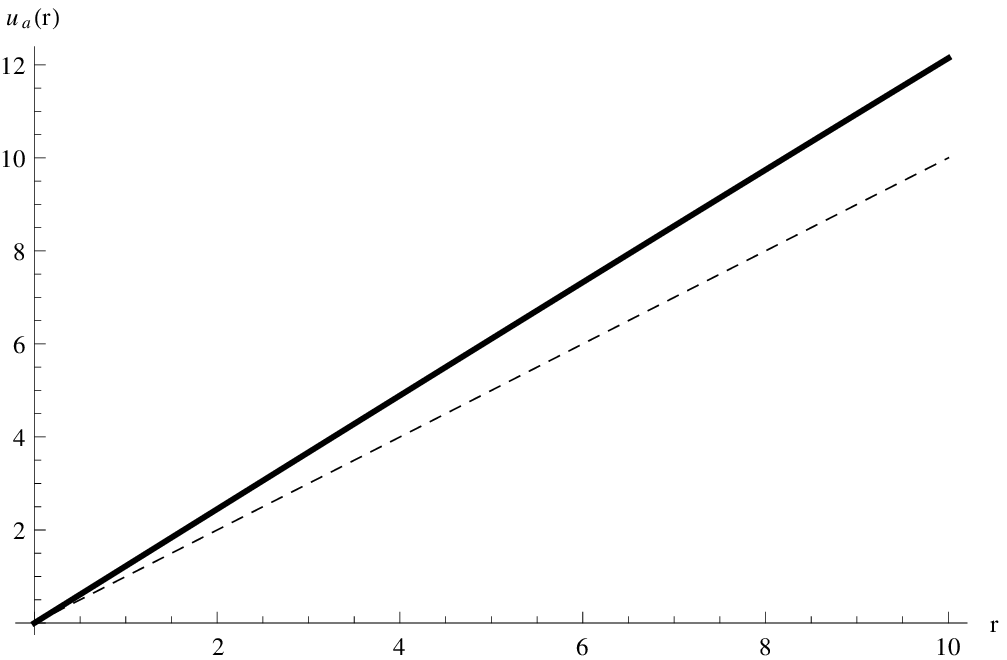}}\\
           \caption{Inverse change of variable. The slope of the graphic for $a=10$ is greater than for $a=0$ (dashed line).}
           \label{changeofvariableinverse}
       \end{minipage}
   \end{figure}

Then, the change of variable has the following form
\begin{displaymath}
 \label{diff}
   r_{a}(u) = \left\{
     \begin{array}{lr}
       u & ,a = 0\\
       -i\sqrt{6}a E\left(\text{arcsinh}\left(\frac{i}{3a}u\right),\frac{3}{2}\right) & , a\neq 0,
     \end{array}
   \right.
\end{displaymath}
where $E$ represents the elliptic integral of second kind. The shape of this change of variable is sketched in Fig. (\ref{variablechange}).


Since $\beta$ is initially defined according to the $u$ variable, it is useful to obtain the inverse change $u:0,\infty) \rightarrow [0,\infty)$ that can be written as
\begin{displaymath}
 \label{inversediff}
   u_{a}(u) = \left\{
     \begin{array}{lr}
       r & ,a = 0\\
       -i\sqrt{6}a E\left(\text{arcsinh}\left(\frac{i}{\sqrt{6}a}r\right),\frac{2}{3}\right) & , a\neq 0,
     \end{array}
   \right.
\end{displaymath}
whose graphic is shown in Fig. (\ref{changeofvariableinverse}).

Henceforward, we shall use the gaussian coordinates $(r,\theta)$ to study the braneworld scenario that we shall describe in next section.

\section{Bulk geometry}
\label{Bulk geometry}
Once defined and studied the geometry of the $2$-cycle of the resolved conifold which we shall use as a transverse space, let us build a six dimensional warped bulk
$\mathcal{M}_{6}$ of form $\mathcal{M}_{6}=\mathcal{M}_{4}\times \mathcal{C}_{2}$, where $\mathcal{C}_{2}$ is the
section of the resolved conifold described in the last section and the $\mathcal{M}_{4}$ is a 3-brane embedded in $\mathcal{M}_{6}$.

The action for the gravitational field is is defined as
\begin{equation}
\label{action}
  S_{g} =\int_{\mathcal{M}_{6}}{\left(\frac{1}{2\kappa_{6}}R-\Lambda +\mathcal{L}_{m}\right)\sqrt{-g}d^{6}x},
\end{equation}
where $\kappa_{6}=\frac{8\pi}{M_{6}^{4}}$, $M_{6}^{4}$ is the six-dimensional bulk Planck mass and $\mathcal{L}_{m}$ is the matter Lagrangian for the source of the geometry. Note that in this convention, the bulk cosmological constant $\Lambda$
has dimension
$[\Lambda]=L^{-6}=M^{6}$.

Consider \textbf{a} static and axisymmetric warped metric between the 3-brane $\mathcal{M}_{4}$ and the transverse resolved conifold $\mathcal{C}_{2}$ given by
\begin{eqnarray}
\label{metricansatz}
ds^{2}_{6} & =  & W(r,c)\hat{g}_{\mu\nu}(x^{\zeta})dx^{\mu}dx^{\nu}+dr^{2} + \gamma(r,c,a) d\theta^{2},
\end{eqnarray}
where
$W:[0,\infty) \rightarrow [0,\infty)$, $W\in C^{\infty}$ is the so called warp function. Henceforward, we shall use the following warp function firstly proposed in \cite{Silva:2012yj}
\begin{equation}
\label{warpfunction}
W(r,c)=e^{-(cr - \tanh{cr})},
\end{equation}
where, $c\in \mathbb{R}$ whose dimension is $[c]=L^{-1}$.

\begin{figure}[htb] 
       \begin{minipage}[b]{0.48 \linewidth}
           \fbox{\includegraphics[width=\linewidth]{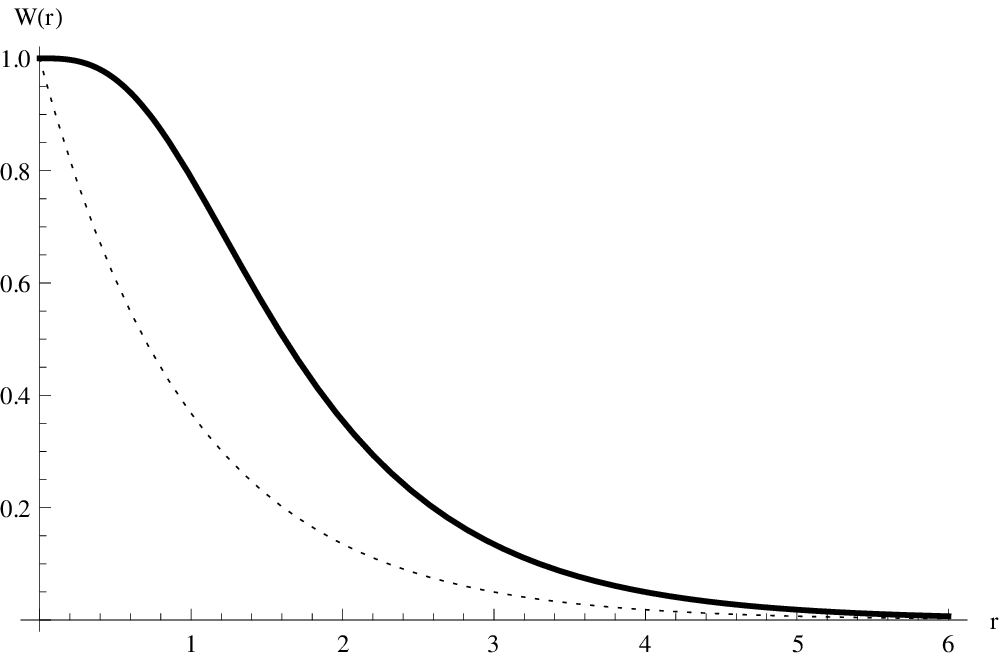}}\\
           \caption{Warp function for $c=1$ (thick line). The thin string warp factor (dotted line) is defined only for the exterior of the string.}
           \label{warp factor}
       \end{minipage}\hfill
       \begin{minipage}[b]{0.48 \linewidth}
           \fbox{\includegraphics[width=\linewidth]{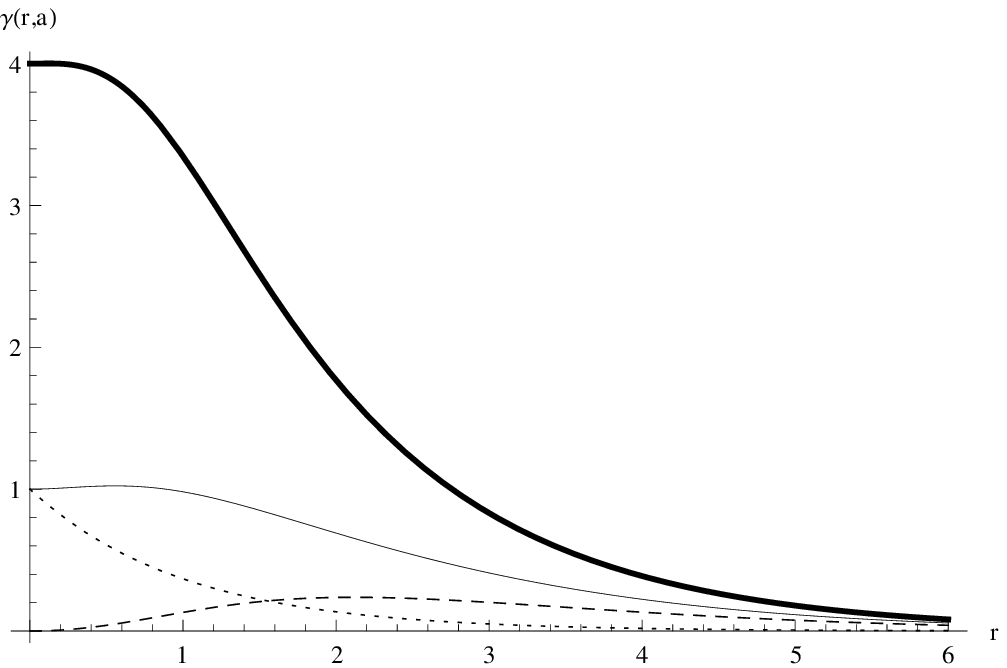}}\\
           \caption{Angular metric component for $c=1$. For $a=0$ (dashed line) there is a conical singularity. For $a=1$ and $a=2$, respectively the thin and thick lines. Dotted line stands for thin string.}
           \label{angularcomponent}
       \end{minipage}
   \end{figure}

It is noteworthy to mention two important features of this warp function. First, as in the usual string-like geometries, the warp function vanishes asymptotically
\cite{Cohen:1999ia,Gherghetta:2000qi,Giovannini:2001hh,Gregory:1999gv,Liu:2007gk,Oda:2000zc,Olasagasti:2000gx,Tinyakov:2001jt}. Second, the warp function chosen above, unlike the thin string-like
geometries
\cite{Gherghetta:2000qi,Oda:2000zc}, satisfies the following conditions for regularity in the origin
\begin{eqnarray}
 W(0,c)=1 & , & W'(0,c)= 0,
\end{eqnarray}
where, the prime $(')$ stands for the derivative $\frac{d}{dr}$. This feature is due to the addition of the term $\tanh{c r}$ that smooth the warp factor near the origin and converges to the
string-like one for large $r$.
Indeed, the warp function has a bell-shape as sketched in the Fig. $(\ref{warp factor})$. Therefore, we can realize this warp function as a near brane correction to the thin string-like
models \cite{Gherghetta:2000qi,Liu:2007gk,Oda:2000zc,Olasagasti:2000gx}. For the thin models (represented in Fig. $(\ref{warp factor})$ by a dotted line) this warp factor function is defined for
the exterior of string only and it can only be regard as defined for all $r$ whether the width of the core of the string-like brane is zero. For $c=1$ the warp factor is presented in
Fig. $(\ref{warp factor})$ by a thick line. Besides, its derivative does not vanishes at the origin as required.

For the angular metric component, $\gamma:[0,\infty) \rightarrow [0,\infty)$, we have chosen the following ansatz
\begin{eqnarray}
\label{angularmetric}
\gamma(r,c,a) & = & W(r,c)\beta(r,a)\nonumber\\
              & = & e^{-(cr - \tanh{cr})}\left(\frac{u(r,a)^{2}+6a^{2}}{6}\right).
\end{eqnarray}
As in the thin string models, the angular metric component vanishes at infinity \cite{Gherghetta:2000qi,Liu:2007gk,Oda:2000zc,Olasagasti:2000gx}. However, a new feature occurs at the origin, where
\begin{equation}
\gamma(0,c,a)=a^{2}.
\end{equation}
For a string-like geometry it is usually assumed the regularity condition \cite{Gherghetta:2000qi,Tinyakov:2001jt,Bostock:2003cv,Kanno:2004nr}

\begin{equation}
\label{regularitycondition}
\gamma(0)=0,
\end{equation}
which is not satisfied for the exterior string solution \cite{Gherghetta:2000qi,Oda:2000zc}

\begin{equation}
\gamma(r)=R_{0}e^{-cr}.
\end{equation}

Note that for $r=0$ there is a 4-brane $\mathcal{M}_{5}$ whose metric is given by
\begin{equation}
ds^{2}_{5}=\hat{g}_{\mu\nu}(x^{\zeta})dx^{\mu}dx^{\nu}+ a^{2} d\theta^{2}.
\end{equation}

Therefore, $\mathcal{M}_{4}\subset \mathcal{M}_{5}=\mathcal{M}_{4}\times S^{1}_{a}$, where $S^{1}_{a}$ is the circle of radius $a$ and $\mathcal{M}_{4}$ is obtained for $a=0$. Consequently,
the geometrical flow of the resolved conifold leads to a dimensional reduction $\mathcal{M}_{6}\rightarrow \mathcal{M}_{5}$ in the origin. The string-like dimensional reduction $\mathcal{M}_{6}\rightarrow \mathcal{M}_{4}$ is reached provided $a=0$. In the case $(a=0)$, the condition (\ref{regularitycondition}) is satisfied, since $\gamma(r,c,0)=\frac{r^{2}}{6}$. Yet, as we will show in the sections (\ref{masslessmode}) and (\ref{quantumpotential}), the gauge field is ill-defined if condition (\ref{regularitycondition}) is accepted (conical space).

For the brane at resolved conifold, the angular component has a $Z_{2}$ symmetry due to its zero derivative at origin. In Fig. (\ref{angularcomponent}), for $a=0$ (dashed line), the component exhibits a conical behavior at
the origin, which is expected for a string-like geometry. For $a=1$ and $a=2$, respectively the thin and thick lines, the function acquires a bell-shape.

The scalar curvature of $\mathcal{M}_{6}$ is given by
\begin{eqnarray}
 R & = &
\frac{\hat{R}}{\sigma}-\Big[4\left(\frac{W'}{W}\right)' + \left(\frac{W'}{W}\right)^{2} + \left(\frac{\gamma'}{\gamma}\right)' - \frac{1}{2}
\left(\frac{\gamma'}{\gamma}\right)^{2} + 2\frac{W'}{W}\frac{\gamma'}{\gamma} \Big],
\end{eqnarray}
where, $\hat{R}$ is the scalar curvature of the 3-brane $\mathcal{M}_{4}$.

\begin{figure}
 \centering
 \includegraphics[scale=1.1]{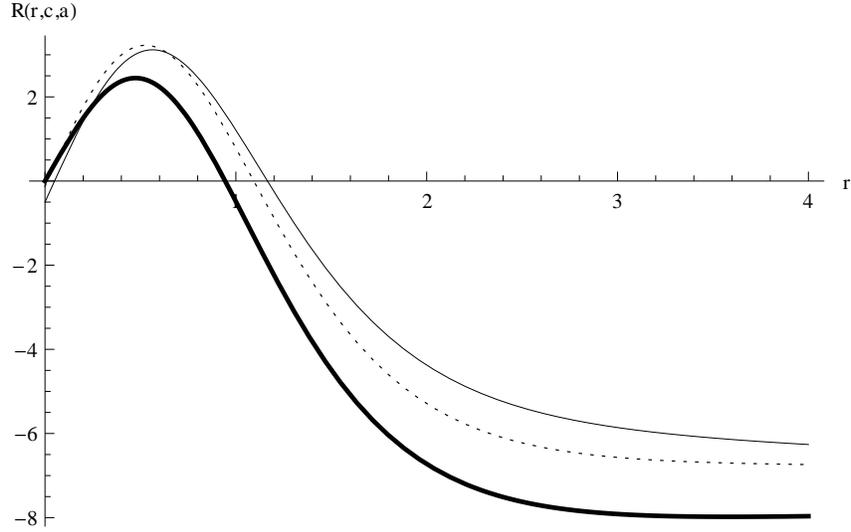}
\caption{Bulk scalar curvature for $\hat{R}=0$ and $c=1$. The manifold is smooth everywhere and it approaches asymptotically to an $AdS_{6}$.}
\label{scalarcurvature}
\end{figure}

The scalar curvature is plotted in the Fig. (\ref{scalarcurvature}) for $\hat{R}=0$. Note that the manifold has a smooth geometry everywhere and it approaches asymptotically to an
$AdS_{6}$  which radius does not depend on the resolution parameter. This ensure us to claim that $(\mathcal{M}_{6},ds^{2}_{6})$ is an extension of the thin string models for both near and far from the brane.

The properties of the scenario described above turned it an interesting extension of the string-like braneworld. Indeed, it not only allows to regularize the geometry at the origin as it yields a
geometric flow in the transverse space that changes the brane properties, as we shall see later.

In this work, we shall not deduce this geometric solution from a matter Lagrangian. This would demand numerical analysis, as performed in (\cite{Giovannini:2001hh}) that diverges of our aim. Instead, following the same approach used in (\cite{Gherghetta:2000qi,Tinyakov:2001jt,Oda2000a}), our main goal is to study the behavior of the gauge field minimally coupled in this background when the geometry suffers a resolution flow. In order to investigate the physical feasibility of this model, in the next section we shall study
the properties of the stress-energy-momentum tensor and the string tensions as well.

\subsection{Einstein equation}
\label{einsteinequation}
Through the Einstein equations, in this section we shall derive and study some physical quantities of this scenario, as the components of stress-energy-momentum tensor, the value of the cosmological
constant, the string tensions and the relationship between the mass scales.

Firstly, let us assume a cylindrically symmetric ansatz for the energy-momentum tensor
\begin{align}
\label{energymomentumansatz}
 T^{\mu}_{\nu} & = t_{0}(r)\delta^{\mu}_{\nu},\\
 T^{r}_{r} & = t_{r}(r),\\
 T^{\theta}_{\theta} & = t_{\theta}(r),
\end{align}
where,
\begin{equation}
T_{ab}=\frac{2}{\sqrt{-g}}\frac{\partial \mathcal{L}_{m}}{\partial g^{ab}}.
\end{equation}

From the action (\ref{action}) we obtain the Einstein equations
 \begin{equation}
\label{Einstein}
 R_{ab}-\frac{R}{2}g_{ab} = -\kappa_{6}(\Lambda g_{ab} + T_{ab}).
\end{equation}

The metric ansatz (\ref{metricansatz}) leads the Einstein equations to a system of coupled ordinary differential equations, namely
\begin{eqnarray}
2\left(\frac{W'}{W}\right)' + \frac{5}{2}\left(\frac{W'}{W}\right)^2 + \frac{5}{4}\frac{W'}{W}\frac{\beta'}{\beta} + \frac{1}{2}\left(\frac{\beta'}{\beta}\right)'+ \frac{1}{4}\left(\frac{\beta'}{\beta}\right)^2 & = & -\kappa_{6}(\Lambda + t_{0}) + \kappa_{4}\frac{\Lambda_{4}}{W}\\
\frac{5}{2}\left(\frac{W'}{W}\right)^2 + \frac{W'}{W}\frac{\beta'}{\beta} & = & -\kappa_{6}(\Lambda + t_{r}) + 2\kappa_{4}\frac{\Lambda_{4}}{W}\\
2\left(\frac{W'}{W}\right)' + \frac{5}{2}\left(\frac{W'}{W}\right)^2 & = & -\kappa_{6}(\Lambda + t_{\theta}) + 2\kappa_{4}\frac{\Lambda_{4}}{W},
\label{einsteinequations}
\end{eqnarray}
where, we assumed the 4-brane $\mathcal{M}_{4}$ to be a maximally symmetric manifold whose four-dimensional cosmological constant on the 3-brane satisfies
\begin{equation}
 \hat{R}_{\mu\nu}-\frac{\hat{R}\hat{g_{\mu\nu}}}{2}=\kappa_{4}\Lambda_{4}\hat{g}_{\mu\nu},
\end{equation}
with, $\kappa_{4}=\frac{8\pi}{M_{4}^{2}}$.
However, since we have chosen a metric ansatz where $\lim_{r\rightarrow\infty}{W(r)=0}$, the last term in Einstein equations blows up at infinity. In order to overcome this obstacle, hereinafter we will set $\Lambda_{4}=0$, i.e., $\mathcal{M}_{4}$ is a flat-brane.

Using the metric ansatz (\ref{metricansatz}), we have found the following components of the energy-momentum tensor
\begin{displaymath}
 \label{tr}
   t_{0}(r,c,a) = \left\{
     \begin{array}{lr}
        \frac{1}{\kappa_{6}}\Big[c^{2}\left(5\text{sech}^{2}cr + 4\text{sech}^{2}cr \tanh{cr} -\frac{5}{2}\text{sech}^{4} c r\right) + \\
           + \frac{5c}{2}\frac{\tanh{c r}^{2}}{r} & ,a = 0\\
        \frac{1}{\kappa_{6}}\Big[c^{2}\left(5\text{sech}^{2}cr + 4\text{sech}^{2}cr \tanh{cr} -\frac{5}{2}\text{sech}^{4} c r\right) + \\
           + \frac{5}{2}\frac{u}{u^{2} +6a^{2}}\tanh{c r}^{2}\sqrt{\frac{r^{2}+9a^{2}}{r^{2}+6a^{2}}}+3a^{2}\frac{u}{u^{2} +6a^{2}}\frac{r}{(r^{2}+6a^{2})^{\frac{3}{2}}(r^{2}+9a^{2})^{\frac{1}{2}}} &,  a\neq 0,
     \end{array}
   \right.
\end{displaymath}
\begin{displaymath}
 \label{tr}
   t_{r}(r,c,a) = \left\{
     \begin{array}{lr}
       \frac{1}{\kappa_{6}}\Big[c^{2}\Big(5\text{sech}^{2}cr -\frac{5}{2}\text{sech}^{4} c r\Big) + c\Big(\tanh^{2}{cr}\frac{2}{r}\Big)\Big] & ,a = 0\\
       \frac{1}{\kappa_{6}}\Big[c^{2}\Big(5\text{sech}^{2}cr -\frac{5}{2}\text{sech}^{4} c r\Big) + c\Big(\tanh^{2}{cr}\frac{2u}{u^{2}+6a^{2}}\sqrt{\frac{r^{2}+9a^{2}}{r^{2}+6a^{2}}}\Big)\Big] & , a\neq 0,
     \end{array}
   \right.
\end{displaymath}
\begin{equation}
 t_{\theta}(r,c)  = \frac{c^2}{\kappa_{6}}\Big(5\text{sech}^{2}cr + 4\text{sech}^{2}cr \tanh{cr} -\frac{5}{2}\text{sech}^{4} c r\Big).
\end{equation}

\begin{figure}[htb] 
       \begin{minipage}[b]{0.48 \linewidth}
           \fbox{\includegraphics[width=\linewidth]{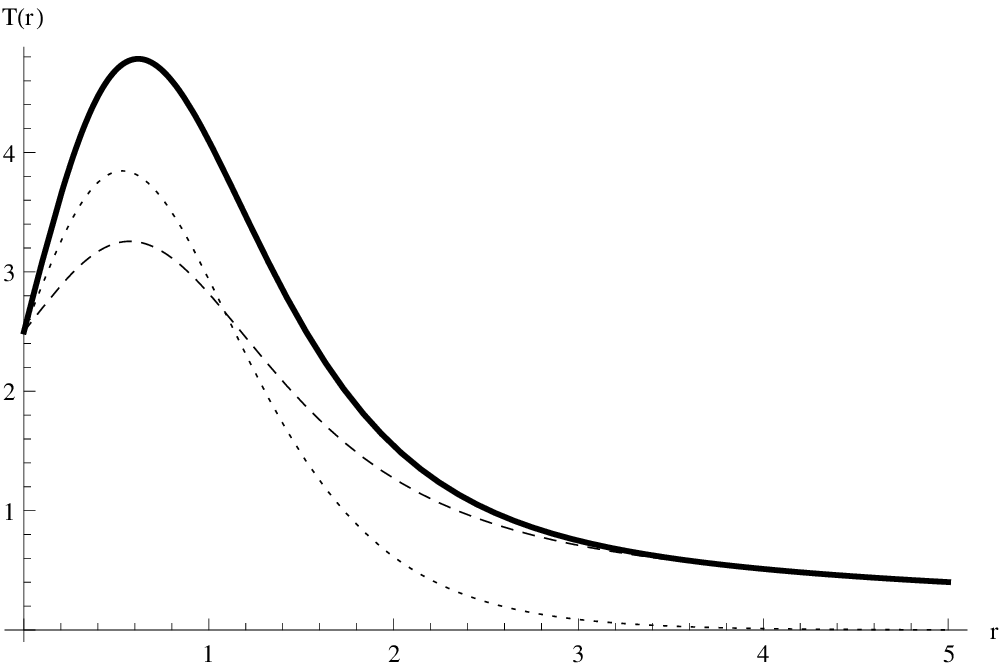}}\\
           \caption{Components of energy-momentum tensor for $a=0$. The energy density (thick line) satisfies the dominant energy condition since it is greater than or equals to the radial (dashed line) and angular component (dotted line).}
           \label{energymomentum0}
       \end{minipage}\hfill
       \begin{minipage}[b]{0.48 \linewidth}
           \fbox{\includegraphics[width=\linewidth]{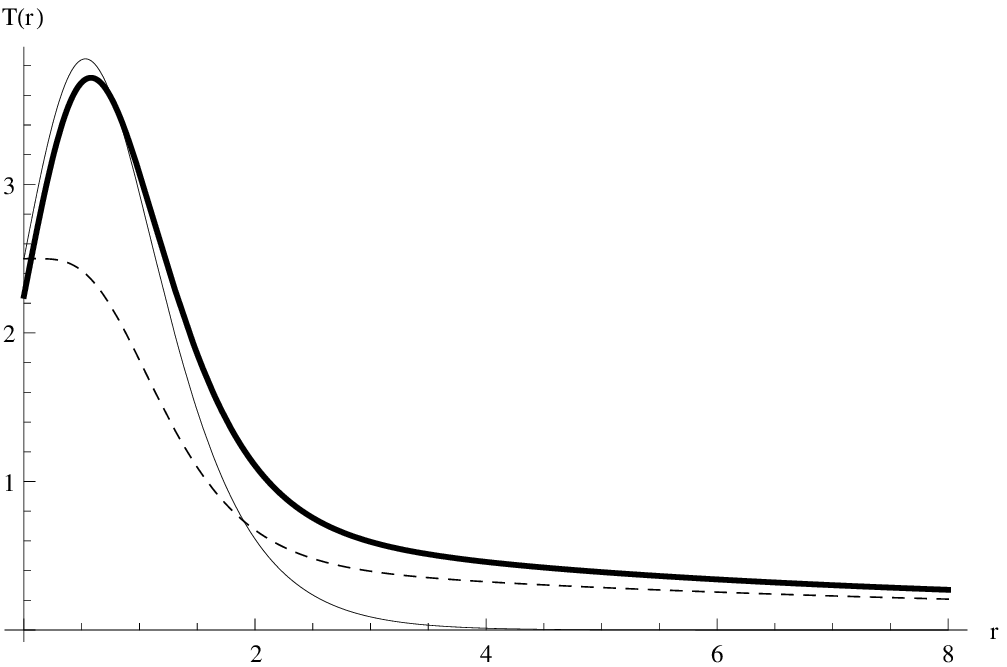}}\\
           \caption{Components of energy-momentum tensor for $a=1$. All the components satisfy the weak energy condition and the energy density (thick line) and the angular pressure (thin line) are quite similar.}
           \label{energymomentum1}
       \end{minipage}
\end{figure}

\begin{figure}[htb] 
       \begin{minipage}[b]{0.48 \linewidth}
           \fbox{\includegraphics[width=\linewidth]{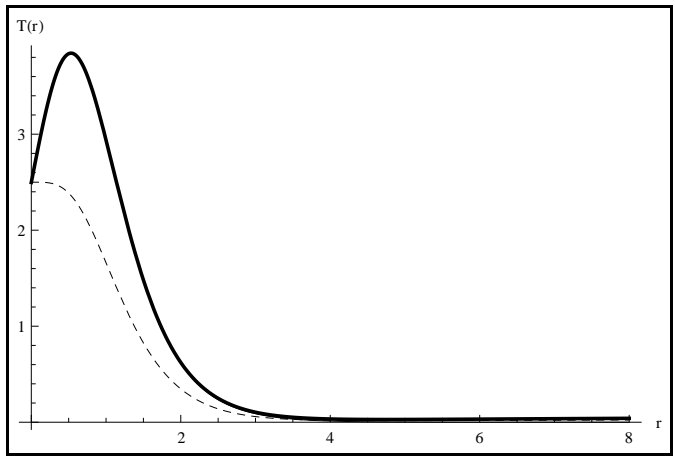}}\\
           \caption{Components of energy-momentum tensor for $a=10$. Note that for a high value of $a$ the energy density (thick line) and the angular pressure (dashed line) converge quickly to the same value.}
           \label{energymomentum10}
       \end{minipage}\hfill
\end{figure}

The graphics of these functions were plotted in Figures (\ref{energymomentum0}), (\ref{energymomentum1}), and (\ref{energymomentum10}).
It is worthwhile to say that all the components have compact support near the origin
where the 3-brane is. This feature also appears in a string-like brane generated by a vortex \cite{Giovannini:2001hh}. Hence, the source of this geometry, whatever it be, has its energy content localized near the origin, and so the geometry is created by a local source.

Furthermore, the components satisfy weak energy conditions, unless for very tiny values of $a$. For $a=0$, the energy conditions (weak, strong and dominant) are all fulfilled. These features
turn this geometry a physical and realistic scenario, albeit exotic, since the components do
not satisfy the dominant and the strong energy conditions for all values of $a$. Besides, the dominant energy condition is also broken in the Gherghetta-Schaposhnikov (GS)  model (infinitely thin string) \cite{Gherghetta:2000qi,Tinyakov:2001jt}.

For large values of $a$, the energy density and the angular pressure converge equally for the same value and both are always greater than the radial pressure. Besides, since
all components decay quickly we can estimate a width of the brane and it decreases as $a$ increases.


Another consequence of this geometry is that the bulk has a negative cosmological constant. Indeed, for large $r$, the components of energy momentum tensor vanish and the vacuum
solution of the Einstein equations yields the well-known relationship \cite{Gherghetta:2000qi,Oda:2000zc}:
\begin{equation}
 c^{2}=-\frac{2K_{6}}{5}\Lambda.
 \label{cosmologicalconstantrelation}
\end{equation}


\subsection{String tensions}
\label{stringtensions}

An important quantity of the brane is the so-called the brane tension.
We define the 4-tension per unit of volume of the 3-brane, as \cite{Gherghetta:2000qi,Oda:2000zc,Giovannini:2001hh}

\begin{eqnarray}
  \mu_{i}(c,a) & = & \int_{0}^{\infty}{t_{i}(r,c,a)W^{2}(r,c)\sqrt{\gamma(r,c,a)}dr}
\end{eqnarray}
Note that $[\mu_{i}]=M^{4}$. The energy string tension is sometimes called the mass \textit{per} unit of volume of the string. It is useful to define the tensions \textit{per} unit of
length by means of a smooth function $\Xi : \mathbb{R}^{3}\rightarrow \mathbb{R}$ given by

\begin{equation}
 \Xi_{i}(r,c,a)=t_{i}(r,c,a)W^{2}(r,c)\sqrt{\gamma(r,c,a)}.
\end{equation}

\begin{figure}
       \begin{minipage}[b]{0.48 \linewidth}
           \fbox{\includegraphics[width=\linewidth]{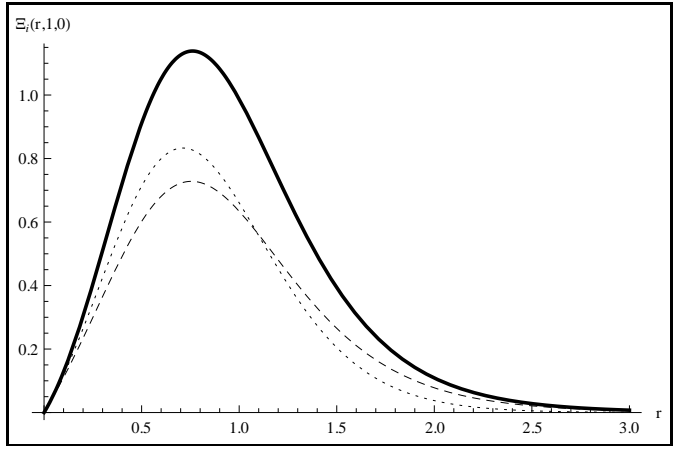}}\\
           \caption{String tensions \textit{per} unit of length for $a=0$. The energy string tension (thick line) is greater than the angular string tension (dotted line) and the radial string tension (dashed line).}
           \label{stringtension_0}
       \end{minipage}
\end{figure}

\begin{figure}[htb] 
       \begin{minipage}[b]{0.48 \linewidth}
           \fbox{\includegraphics[width=\linewidth]{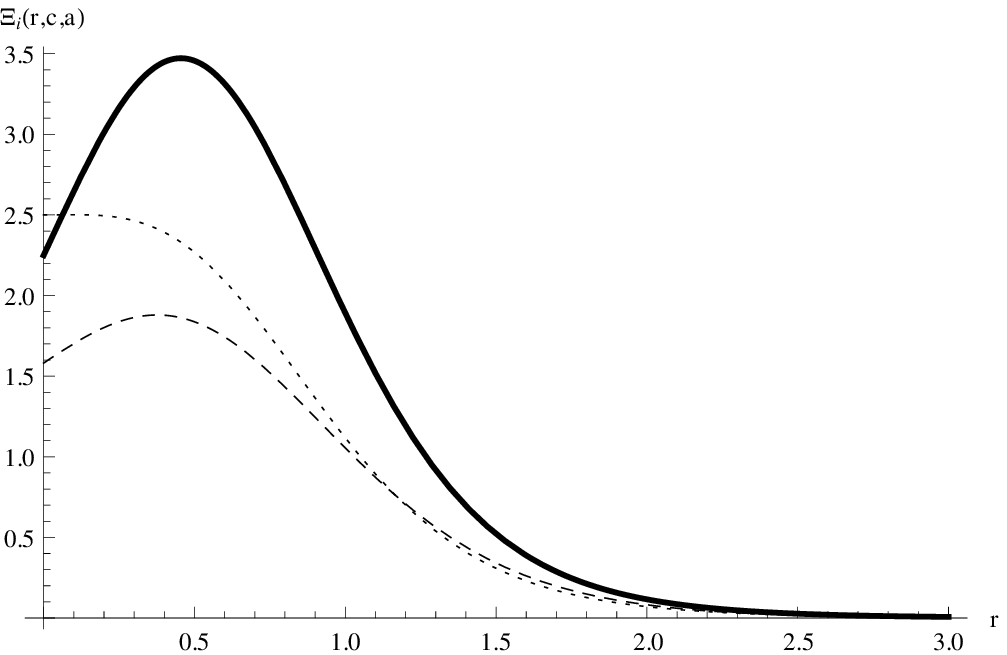}}\\
           \caption{String tensions \textit{per} unit of length for $a=1$. The difference between the angular and energy tension decreases.}
           \label{stringtension_1}
       \end{minipage}\hfill
       \begin{minipage}[b]{0.48 \linewidth}
           \fbox{\includegraphics[width=\linewidth]{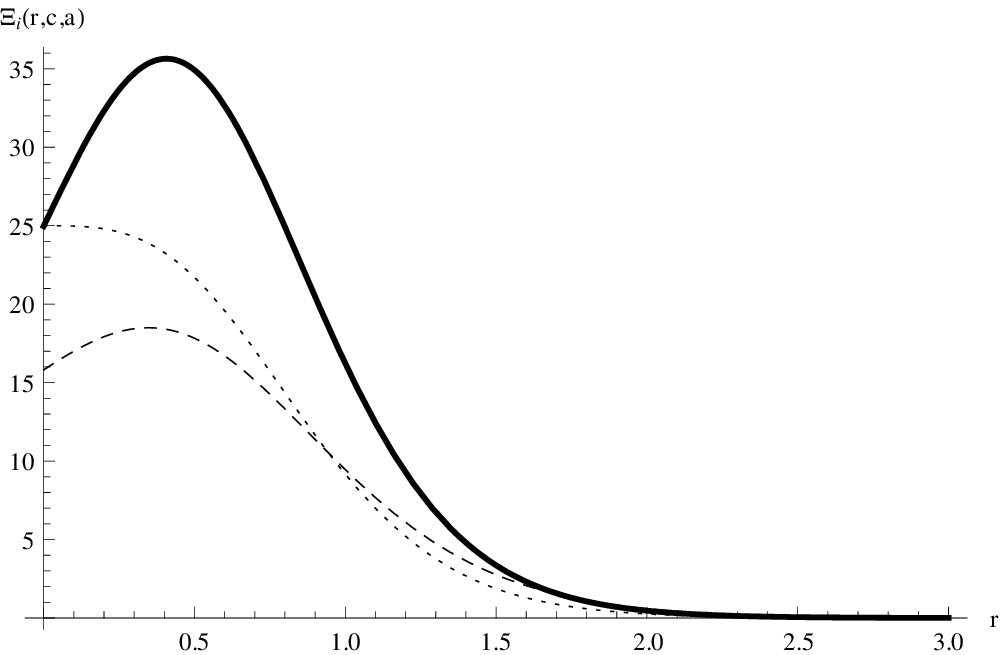}}\\
           \caption{String tensions \textit{per} unit of length for $a=10$. The energy tensions is everywhere greater than other tensions.}
           \label{stringtension_10}
       \end{minipage}
\end{figure}

We have plotted $\Xi_{i}$ for some values of $a$ in the figures (\ref{stringtension_0}),(\ref{stringtension_1}), (\ref{stringtension_10}).
By means of the graphic it is possible to see that for large values of $a$ the energy tension $(\mu_{0})$ is always greater than the radial tension $(\mu_{r})$ and the angular tension $(\mu_{\theta})$. This
result also appears in the other string-like defects \cite{Gherghetta:2000qi,Oda:2000zc,Giovannini:2001hh}. In GS model, for instance, the authors have taken an infinitely thin brane
with $\mu_{r}=0$. Therefore, the differences $\mu_{0}-\mu_{r}$ and $\mu_{0}-\mu_{\theta}$ decrease when $a$ decreases.

\subsection{Mass hierarchy}
\label{masshierarchy}

In the last section, we have seen that string tensions, including the string mass, depend on the evolution parameter $a$. Now we wish to study how the geometric
evolution alters the relation between the bulk and brane mass scale.

In this geometry, the relationship between the four-dimensional Planck mass ($M_{4}$) and the bulk Planck mass
($M_{6}$) is given by \cite{Gherghetta:2000qi,Giovannini:2001hh}
\begin{equation}
\label{planckmass}
M^{2}_{4}=2\pi M_{6}^{4}\int_{0}^{\infty}{W^{2}(r,c)\sqrt{\gamma(r,c,a)}dr}.
\end{equation}

Due to the complexity of the warp factor and the angular metric component, it is a difficult task to find the integral in Eq. $(\ref{planckmass})$. However, we can study it qualitatively. Since all the metric components are limited functions, this geometry has a finite volume and then, it can be used to tuning the ratio between the Planck masses
explaining the hierarchy between them. Note, however, that the relationship between the Planck masses depends on the evolution parameter $a$.

\begin{figure}
\centering
\includegraphics[scale=1.1]{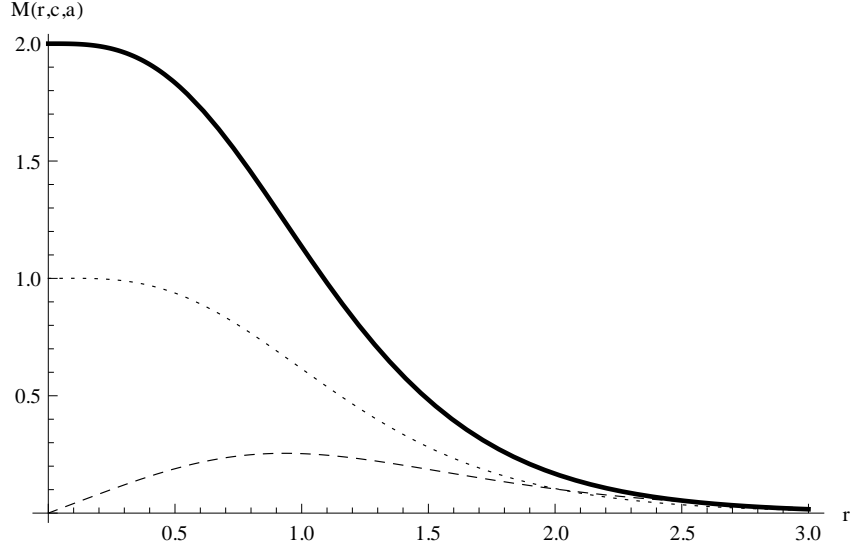}
\caption{M(r,c,a) for some values of $a$ and $c=1$. As $a$ increases ($a=0$ is the dashed line, $a=1$ is the dotted line and $a=2$ is the thick one) M takes higher values and its integral (which gives the relation between the mass scales ) also.}
\label{massrelationship}
\end{figure}

In order to study the behavior of $\frac{M^{2}_{4}}{2\pi M_{6}^{4}}$ we defined the function $M:\mathbb{R}^{3}\rightarrow \mathbb{R}$ given by
\begin{equation}
 M=M(r,c,a)=W^{2}(r,c)\sqrt{\gamma(r,c,a)},
\end{equation}
whose integral over $r$ provides the desired ratio. This function was sketched in Fig. $(\ref{massrelationship})$ where we can conclude that as higher the value of $a$ is, higher is the value of $\frac{M^{2}}{2\pi M_{6}^{4}}$.

Hence, a evolution of the bulk geometry through a geometrical flow, as the resolved conifold, could alters the hierarchy between the fundamental energy scales in the bulk and in the
brane.  This is an extension to the string-like GS model tuning of the Planck masses since in this model the relationship is accomplished by fixed geometrical constants, e.g., the
bulk cosmological constant and the angular tension \cite{Gherghetta:2000qi,Oda:2000zc,Tinyakov:2001jt}. As usual in braneworld models
,
the bulk mass scale is chosen in the
 scale.
 
\section{Gauge vector field localization}
\label{Gauge vector field localization}

Next, we turn our attention to the case of vector field. Let us start with the action of a $U(1)$ vector field
\begin{equation}
\label{gaugefieldaction}
S_{m} =\int{d^{6}x\sqrt{-g}g^{MN}g^{RS}F_{MN}F_{RS}},
\end{equation}
where $F_{MN}=\nabla_{M}A_{N}-\nabla_{M}A_{N}$ as usual. From the action (\ref{gaugefieldaction}) the motion equation is given by
\begin{eqnarray}
\label{gaugefieldmotionequation}
 \frac{1}{\sqrt{-g}}\partial_{R}(\sqrt{-g}g^{RM}g^{LN}F_{MN})= 0.
\end{eqnarray}

In the background metric (\ref{metricansatz}), the equation of motion (\ref{gaugefieldmotionequation}) becomes
\begin{equation}
\label{motionequation1}
\left(\eta^{\mu\nu}\partial_{\mu}\partial_{\nu}+\frac{W(r,c)}{\gamma(r,c,a)}{\partial_{\theta}}^{2}\right)A_{r}=0,
\end{equation}
\begin{equation}
\label{motionequation2}
\partial_{r}\left(\frac{W^{2}(r,c)\sqrt{\gamma(r,c,a)}}{\gamma(r,c,a)}\partial_{\theta}A_{r}\right)=0,
\end{equation}
and
\begin{equation}
\label{equationmotion3}
\Big(\eta^{\mu\nu}\partial{\mu}\partial{\nu}+\frac{W}{\gamma}\partial_{\theta}^{2}+\frac{1}{\sqrt{\gamma}}\partial_{r}W\sqrt{\gamma}\partial_{r}\Big) A_{\lambda}
-\frac{1}{\sqrt{\gamma}}\partial_{r}(W\sqrt{\gamma}\partial_{\lambda}A_{r})=0,
\end{equation}
where we have used the usual gauge conditions \cite{Oda2000a,Oda:2000zc}
\begin{equation}
\partial_{\mu}A^{\mu}=A_{\theta}=0.
\end{equation}
Let us take the following forms of the $KK$ decomposition as usual
\begin{equation}
\label{equationmotion4}
A_{\mu}(x^M)=\sum_{l=0}^{\infty}A_{\mu}^{(l)}(x^{\mu})\chi(r)Y_{l}(\theta),
\end{equation}
and
\begin{equation}
\label{equationmotion5}
A_{r}(x^M)=\sum_{l=0}^{\infty}A_{r}^{(l)}(x^{\mu})\xi(r)Y_{l}(\theta).
\end{equation}

Then, from  Eq. (\ref{motionequation1}) we have
\begin{equation}
\label{motionequation6}
(\eta^{\mu\nu}\partial_{\mu}\partial_{\nu}- \frac{W}{\gamma}l^2)A_{r}^{(l)}(x^{\mu})=0.
\end{equation}

Therefore Eq. (\ref{motionequation2}) leads to a general solution to $\xi(r)$ as
\begin{equation}
\label{motionequation7}
\xi(r)=\alpha{\gamma}^{1/2}W^{-2},
\end{equation}
with $\alpha$ being \textbf{an} integration constant. The $\xi(r)$ in Eq. $(\ref{motionequation7})$ extends the analog function found by Oda \cite{Oda2000a}. 

Finally, using Eq. (\ref{motionequation7}), we see that Eq. (\ref{equationmotion3}) reduces to the form
\begin{equation}
\label{motionequation8}
\left(m^2-\frac{W}{\gamma}l^2+\frac{1}{\sqrt{\gamma}}\partial_{r}(W\sqrt{\gamma}\partial_{r})\right)A_{\lambda}^{(l)}(x^{\mu})\chi(r)=\frac{1}{\sqrt{\gamma}}
\partial_{\lambda}A_{r}^{(l)}(x^{\mu})\partial_{r}(W\sqrt{\gamma}\xi(r)),
\end{equation}
where we have required $A_{\lambda}^{l}(x^{\mu})$ to satisfy the following relation
\begin{equation}
(\eta^{\mu\nu}\partial_{\mu}\partial_{\nu}- m^2)A_{\lambda}^{l}(x^{\mu})=0.
\end{equation}

The Eq. (\ref{motionequation8}) differs from that presented by Oda in Ref. \cite{Oda2000a} through  the non-homogenous term in the right hand side. This difference arises due to the
function $\xi(r)$ that is more general than that studied in \cite{Oda2000a}.

Let us restrict ourselves to the case where

\begin{equation}
\label{additionalgauge}
\partial_{\lambda}A_{r}^{(l)}(x^{\mu})=0.
\end{equation}
Condition $(\ref{additionalgauge})$ yields the homogeneous differential equation

\begin{equation}
\label{motionequation9}
\left(m^2-\frac{W}{\gamma}l^2+\frac{1}{\sqrt{\gamma}}\partial_{r}(W\sqrt{\gamma}\partial_{r})\right)\chi(r)=0,
\end{equation}

with an additional condition on the gauge field, $A_{r}(x^{\mu},r,\theta)= C \sum_{l=0}{\xi(r)Y_{l}(\theta)}$, where $C\in \mathbb{R}$. Assuming $C=1$, condition $(\ref{additionalgauge})$ means that the radial component has
no dependence on the brane coordinates. In order to keep the condition $(\ref{additionalgauge})$ in agreement with Eq. $(\ref{motionequation6})$, from now on, we shall choose $l=0$ (the $s-$wave state).

In that case the equation $(\ref{motionequation9})$ turns into a Sturm-Liouville one. Further, let us looking for solutions that satisfy the boundary conditions
\cite{Gherghetta:2000qi,Giovannini:2001hh,Oda2000a}

\begin{equation}
\label{boundarycondition}
\chi'(0)=\lim_{r\rightarrow\infty}\chi'(r)=0.
\end{equation}


Naming solutions of Eq. (\ref{motionequation9}) and as $\chi_{i}(r)$ and $\chi_{j}(r)$, the orthogonality relation between them \textbf{is} given by \cite{Gherghetta:2000qi}
\begin{equation}
\label{orthogonalityrelation}
\int_{0}^{\infty}\sqrt{W(r,c)\beta(r,a)}\chi_{i}^{*}\chi_{j}dr=\delta_{ij}.
\end{equation}

Now, we can rewrite Eq. (\ref{motionequation9}) as


 \begin{equation}
 \label{chiequation}
 \chi''(r)+\left(\frac{3}{2}\frac{W'}{W}+\frac{1}{2}\frac{{\beta}'}{\beta}\right)\chi'(r)+\frac{m^2}{W}\chi(r)=0.
 \end{equation}

The analysis of Eq. (\ref{chiequation}) will be done in two stages as follows.
\subsubsection{Massless mode}
\label{masslessmode}

For $m=0$, a constant function $\hat{\xi}_{0}$ is a particular solution of Eq. (\ref{chiequation}). Since the weight function $\sqrt{W(r,c)\beta(r,a)}$ has compact support around the origin, the constant function $\hat{\xi}_{0}$ is a normalizable solution of eq. (\ref{chiequation}). Therefore, from orthogonality relation (\ref{orthogonalityrelation}), we can construct a normalizable
zero-mode solution as \cite{Gherghetta:2000qi,Oda2000a}

\begin{equation}
\label{zeromode}
\chi_{0}(r,a,c)=NW(r,c)^{\frac{1}{2}}{\beta(r,a)}^{\frac{1}{4}},
\end{equation}
where $N$ is a normalization constant given by
\begin{equation}
N^2=\int_{0}^{\infty}{W(r,c)^{\frac{3}{4}}{\beta(r,a)}dr}.
\end{equation}

\begin{figure}
\centering
\includegraphics[scale=1.1]{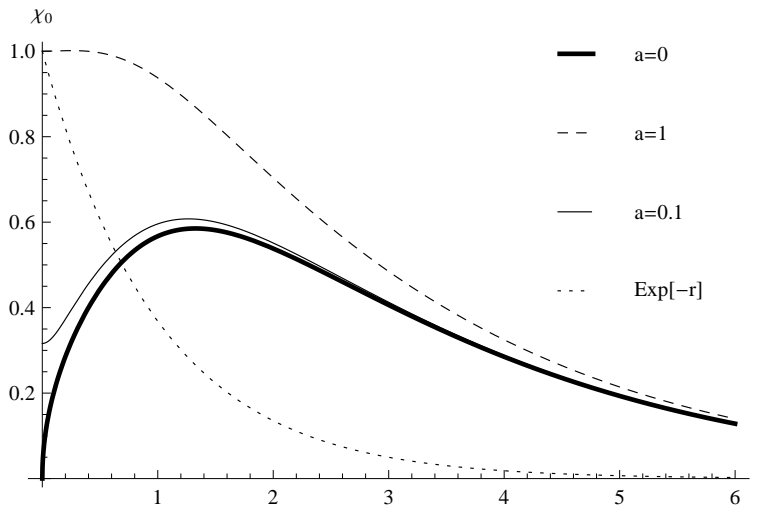}
\caption{$\chi_{0}(r,a)$ for $c=1$ and $a=1$ (dashed line), $a=0.1$ (thin line) and $a=0$ (thick line). Also, the zero mode curve of the string-like defects is displayed for effect of comparison (dotted line).}
\label{ModoZero}
\end{figure}

The massless mode (\ref{zeromode}) is graphically represented in the Fig. (\ref{ModoZero}). As we shall see in section (\ref{quantumpotential}), the zero-mode (\ref{zeromode})
satisfies the analogous Schroedinger equation for $m=0$. Note in figure (\ref{ModoZero}) that the conifold parameter smoothes the zero-mode at the origin. For $a=0$ (thick line), the massless mode vanishes at the origin. This means that, for a string-like geometry satisfying the regularity condition $\gamma(0)=0$ (\ref{regularitycondition}), the vector zero-mode is ill-defined. This issue does not appear in the usual string-like models (dotted line) because these geometries are only exterior string solutions, i.e., $\beta$ is constant \cite{Oda:2000zc}. For $a=0.1$ (thin), the zero-mode is smoothed and shows
an increasing conical behavior near the origin.

This smooth behavior provides a $Z_{2}$ symmetry to the zero mode. Besides, all curves are smooth and satisfy the boundary condition at the origin, in contrast with the solutions found in thin brane models. Note also that the exponentially decreasing displayed on the plot is equal to the one found in the string like defects \cite{Oda:2000zc,Oda2000a}.

\subsubsection{Massive modes}

Now let us study the properties of Eq. (\ref{chiequation}) for $m\neq 0$. Using the expressions for the metric factor we obtain
\begin{equation}
\label{CompleteEquationChi}
\chi''+\left(-\frac{3}{2}c{\tanh(cr)}^2+\frac{u(u^2+9a^2)}{(u^2+6a^2)^{3/2}}\right){\chi}'+\exp(cr-\tanh{cr})m^2\chi=0.
\end{equation}

For $r\rightarrow\infty$, Eq. (\ref{CompleteEquationChi}) converges to the well-known equation of string-like defects given by
\begin{equation}
\label{CompleteEquationFar}
\chi''-\frac{3}{2}c{\chi}'+m^2e^{cr-1}\chi=0,
\end{equation}
and then we argue that asymptotically the behavior of the field has the same features as in string-like defects with a mass term shift $m\rightarrow e^{-1/2}m$ \cite{Oda2000a,Oda:2000zc}.
The solution of Eq. (\ref{CompleteEquationFar}) can be written in terms of the the Bessel function as \cite{Oda2000a,Oda:2000zc}
\begin{equation}
\chi(r)=e^{\frac{3cr}{4}}\left[C_{1}J_{\frac{3}{2}}\left(\frac{2m}{c}e^{\frac{cr-1}{2}}\right)+C_{2}Y_{\frac{3}{2}}\left(\frac{2m}{c}e^{\frac{cr-1}{2}}\right)\right],
\end{equation}
hence, the Kaluza-Klein spectrum  corresponds to the string-like one.

In the near brane regime, for $r\rightarrow{0}$ and $a\neq 0$, Eq. (\ref{CompleteEquationChi}) turns to be
\begin{equation}
\label{CompleteEquationNear}
\chi''+\frac{r}{4a^2}{\chi}'+m^2\chi=0.
\end{equation}
The solution to the Eq. $(\ref{CompleteEquationNear})$ can be written in terms of the Hermite polynomials and the Kummer's function of the first kind of order

\begin{equation}
\label{order}
 n=-1+4a^2m^2,
\end{equation}
as
\begin{equation}
\label{massivemodenear}
\chi(r)=e^{-\frac{r^2}{8a^2}}\left[C_{1}H_{n}\left(\frac{r}{2{\sqrt2}a}\right)+C_{2}{_1F_1}\left(-\frac{1}{2}n,\frac{1}{2},\frac{r^2}{8a^2}\right)\right].
\end{equation}

\begin{figure}
\centering
\includegraphics[scale=1.1]{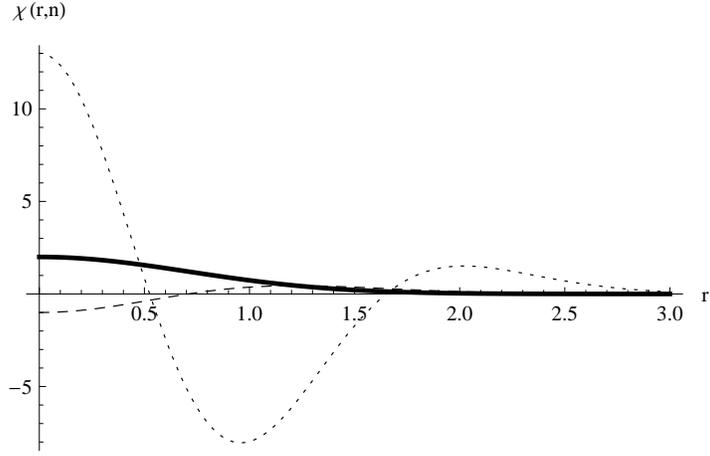}
\caption{$\chi(r,n)$ for some values of $n$, $n=0$ (thick line), $n=2$ (dashed line) and $n=4$ (dotted line).}
\label{Solucaoproxima}
\end{figure}

The solutions near the brane for some $n$ values are displayed in Fig. $(\ref{Solucaoproxima})$ where it is possible to see that the KK modes are
well-defined and smooth functions. Therefore, we claim the resolution parameter alters slightly the KK modes near the brane and provides the same KK modes
as for the exterior string-like model \cite{Oda:2000zc}.

\subsubsection{Quantum potential}
\label{quantumpotential}

Another way to study the massive modes lies on transform the Eq. (\ref{chiequation}) into a Schroedinger-like equation and study its quantum potential. In order to do this, let us make the following change of variable $r\Rightarrow z=f(r)$ such that
\begin{equation}
\frac{dz}{dr}= \upsilon(r).
\end{equation}

This give us the following equation
 \begin{equation}
 \label{FirstDotEquation}
\ddot{\chi}+\left(\frac{3}{2}\frac{\dot{W}}{W}+\frac{1}{2}\frac{\dot{\beta}}{\beta}+\frac{\dot{\upsilon}}{\upsilon}\right)+m^2\frac{\chi}{{\upsilon}^2 W}=0,
\end{equation}
where the dot means derivative with respect to $z$. Choosing ${\upsilon}^2W=1$, we obtain
\begin{equation}
\label{changevariable}
z=z(r)=\int^{r}W^{-1/2}dr'.
\end{equation}

Using the change of variable (\ref{changevariable}) in the equation (\ref{FirstDotEquation}), we get
\begin{equation}
\label{SecondDotEquation}
\ddot{\chi}(z)+\left(\frac{\dot{W}}{W}+\frac{1}{2}\frac{\dot{\beta}}{\beta}\right)\dot{\chi}(z)+m^2\chi(z)=0.
\end{equation}

In order to simplify the last equation, let us write $\chi(z)$ in the form
\begin{equation}
\chi(z)=\Omega(z)\Psi(z)
\end{equation}

This change transform the Eq. (\ref{SecondDotEquation}) to the form
\begin{equation}
\ddot{\Psi}+\left(2\frac{\dot{\Omega}}{\Omega}+\frac{\dot{W}}{W}+\frac{1}{2}\frac{\dot{\beta}}{\beta}\right)\dot{\Psi}+\left(m^2+\frac{\ddot{\Omega}}{\Omega}+
\left(\frac{\dot{W}}{W}+\frac{1}{2}\frac{\dot{\beta}}{\beta}\right)\frac{\dot{\Omega}}{\Omega}\right)\Psi=0.
\end{equation}

Making
\begin{equation}
\frac{\dot{\Omega}}{\Omega}=-\frac{1}{2}\left(\frac{\dot{W}}{W}+\frac{1}{2}\frac{\dot{\beta}}{\beta}\right),
\end{equation}
the function $\Psi(z)$ must satisfies
\begin{equation}
\label{schrodingerequation}
-\ddot{\Psi}(z)+V(z)\Psi(z)=m^2\Psi(z),
\end{equation}
where $V(z)$ is given by
\begin{equation}
V(z)=\frac{1}{2}\frac{\ddot{W}}{W}-\frac{1}{4}\left(\frac{\dot{W}}{W}\right)^2+\frac{1}{4}\frac{\ddot{\beta}}{\beta}-
\frac{3}{16}\left(\frac{\dot{\beta}}{\beta}\right)^2 + \frac{1}{4}\frac{\dot{\beta}}{\beta}\frac{\dot{W}}{W}.
\end{equation}

It is still possible to write the potential in terms of the $ r $ variable. In this case we have
\begin{equation}
V(r,c,a)=\frac{1}{4}W\left[\left(\frac{W'}{W}+\frac{1}{2}\frac{\beta'}{\beta}\right)^2+
2\left(\frac{W'}{W}+\frac{1}{2}\frac{\beta'}{\beta}\right)'+\frac{W'}{W}\left(\frac{W'}{W}+\frac{1}{2}\frac{\beta'}{\beta}\right)\right].
\end{equation}

\begin{figure}
\centering
\includegraphics[scale=1.1]{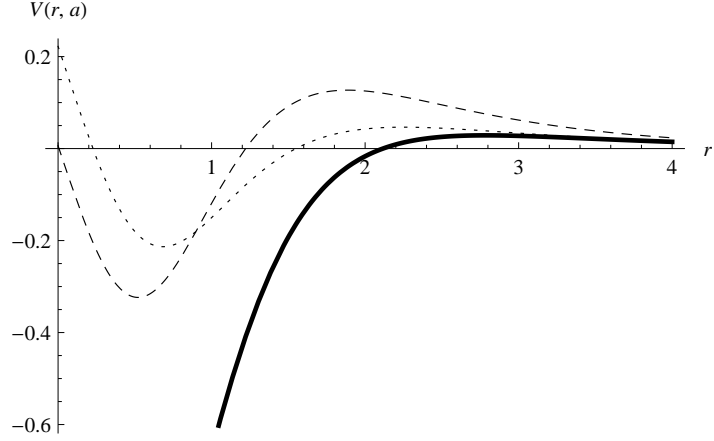}
\caption{$V(r,c,a)$ for some values of $a$ and $c=1$. For $a=0$ (thick line), $a=0.5$ (dotted line), $a=3$ (dashed line).}
\label{PotencialQuantico}
\end{figure}

The Schroedinger-type potential is plotted in Fig. $(\ref{PotencialQuantico})$. It is worthwhile to mention that the massless mode (\ref{zeromode})
satisfies the Schroedinger equation (\ref{schrodingerequation}) for $m=0$, as required \cite{Liu:2011ysa,Liu:2012gv,Liu:2007gk}. The resolution parameter controls the high and depth of the potential well and the value of the potential at the origin.

Moreover, note that for $a=0$ the potential diverges at the origin yielding tachyonic modes on the brane. On the other hand, for $a\neq 0$ the potential is positive at the origin. Then, the resolution parameter also allows to rule out tachyonic vector modes on the brane.

\section{Conclusions and perspectives}
\label{Conclusions and perspectives}

In this work we have studied how the geometrical and physical properties of a warped braneworld scenario builded from a $3$-brane and a $2$-cycle of the resolved conifold change when this manifold evolves under the resolution flow. Then, we analyzed the behavior of the vector gauge  field in this axysymmetric and static six dimensional space-time.

Firstly, we have shown that this is a physical feasible scenario, since the weak energy condition is satisfied for all values of the resolution parameter $a$ and for $a=0$, the source obeys the dominant energy condition.
Moreover, since the transverse manifold has finite volume with negative cosmological constant that does not depend on the resolution parameter and in addition, the bulk is everywhere smooth, this geometry can be regarded as a near
brane extension of the well-known thin string-like brane solutions.

The string tension conditions of the thin string branes are also fulfilled, albeit the radial tension is not zero. Further, the relation between the bulk and brane scale energy increases when the parameter increases.

In the analysis of $s-$wave state of the gauge vector field in this scenario we obtained a normalizable massless mode and a Kaluza-Klein modes near the brane. We found that the resolution parameter avoids the zero mode to vanish on the brane (what happen whether $\beta(0)=0$) and provides a $Z_{2}$ symmetry to this mode. Thus, we argue that resolution parameter smoothes the vector modes near the brane and it agrees with the string-like zero-mode for large distances.

Moreover, the zero-mode satisfies the Schroedinger-like equation for $m=0$, as required. The potential has an infinite potential well for $a=0$
what leads to tachyonic modes on the brane. For $a\neq0$, the potential is positive and repulsive at the origin. The massive modes are smooth near the brane and have the same behavior of the string-like model asymptotically. Then, the resolution parameter allows the existence of a well-defined vector
zero-mode and avoids KK and tachyonic modes on the brane.

For future works we intend to study the behaviour of the gauge vector field for $l\neq 0$, and to study other fields in this background as well.

\section{Acknowledgments}

Financial support from the Brazilian agencies CNPq and CAPES is gratefully acknowledged.


\begin{thebibliography}{99}

\bibitem{Randall:1999vf}
  L.~Randall and R.~Sundrum,
  Phys.\ Rev.\ Lett.\  {\bf 83}, 4690 (1999)
  [arXiv:hep-th/9906064].


\bibitem{Randall:1999ee}
  L.~Randall and R.~Sundrum,
  Phys.\ Rev.\ Lett.\  {\bf 83}, 3370 (1999)
  [arXiv:hep-ph/9905221].

\bibitem{Goldberger:1999uk}
  W.~D.~Goldberger and M.~B.~Wise,
  Phys.\ Rev.\ Lett.\  {\bf 83}, 4922 (1999)
  [hep-ph/9907447].


\bibitem{Csaki:2000fc}
  C.~Csaki, J.~Erlich, T.~J.~Hollowood and Y.~Shirman,
  Nucl.\ Phys.\ B {\bf 581}, 309 (2000)
  [hep-th/0001033].


\bibitem{Gremm:1999pj}
  M.~Gremm,
  Phys.\ Lett.\ B {\bf 478}, 434 (2000)
  [hep-th/9912060].

\bibitem{Bazeia:2004dh}
  D.~Bazeia and A.~R.~Gomes,
  JHEP {\bf 0405}, 012 (2004)
  [hep-th/0403141].

\bibitem{Almeida:2009jc}
  C.~A.~S.~Almeida, M.~M.~Ferreira, Jr., A.~R.~Gomes and R.~Casana,
  Phys.\ Rev.\ D {\bf 79}, 125022 (2009)
  [arXiv:0901.3543 [hep-th]].

\bibitem{Goldberger:1999wh}
  W.~D.~Goldberger and M.~B.~Wise,
  Phys.\ Rev.\ D {\bf 60}, 107505 (1999)
  [hep-ph/9907218].

\bibitem{Kehagias:2000au}
  A.~Kehagias and K.~Tamvakis,
  Phys.\ Lett.\ B {\bf 504}, 38 (2001)
  [hep-th/0010112].

\bibitem{Huber:2000ie}
  S.~J.~Huber and Q.~Shafi,
  Phys.\ Lett.\ B {\bf 498}, 256 (2001)
  [hep-ph/0010195].


\bibitem{Liu:2011ysa}
  Y.~-X.~Liu, C.~-EFu, H.~Guo and H.~-T.~Li,
  Phys.\ Rev.\ D {\bf 85}, 084023 (2012)
  [arXiv:1102.4500 [hep-th]].

\bibitem{Liu:2012gv}
  Y.~-X.~Liu, F.~-W.~Chen, Heng-Guo and X.~-N.~Zhou,
  JHEP {\bf 1205}, 108 (2012)
  [arXiv:1205.0210 [hep-th]].

\bibitem{Liu:2007gk}
  Y.~-X.~Liu, L.~Zhao and Y.~-S.~Duan,
  JHEP {\bf 0704}, 097 (2007)
  [hep-th/0701010].

\bibitem{Chodos:1999zt}
  A.~Chodos and E.~Poppitz,
  Phys.\ Lett.\ B {\bf 471}, 119 (1999)
  [hep-th/9909199].

\bibitem{Olasagasti:2000gx}
  I.~Olasagasti and A.~Vilenkin,
  Phys.\ Rev.\ D {\bf 62}, 044014 (2000)
  [hep-th/0003300].

\bibitem{Kehagias:2004fb}
  A.~Kehagias,
  Phys.\ Lett.\  B {\bf 600}, 133 (2004)
  [arXiv:hep-th/0406025].

\bibitem{Benson:2001ac}
  K.~Benson and I.~Cho,
  Phys.\ Rev.\ D {\bf 64}, 065026 (2001)
  [hep-th/0104067].


\bibitem{Chen:2000at}
  J.~W.~Chen, M.~A.~Luty and E.~Ponton,
  JHEP {\bf 0009}, 012 (2000)
  [arXiv:hep-th/0003067].


\bibitem{Cohen:1999ia}
  A.~G.~Cohen and D.~B.~Kaplan,
  Phys.\ Lett.\  B {\bf 470}, 52 (1999)
  [arXiv:hep-th/9910132].

\bibitem{Gregory:1999gv}
  R.~Gregory,
  Phys.\ Rev.\ Lett.\  {\bf 84}, 2564 (2000)
  [arXiv:hep-th/9911015].

\bibitem{Gherghetta:2000qi}
  T.~Gherghetta and M.~E.~Shaposhnikov,
  Phys.\ Rev.\ Lett.\  {\bf 85}, 240 (2000)
  [arXiv:hep-th/0004014].

\bibitem{Giovannini:2001hh}
  M.~Giovannini, H.~Meyer and M.~E.~Shaposhnikov,
  Nucl.\ Phys.\ B {\bf 619}, 615 (2001)
  [hep-th/0104118].


\bibitem{Oda:2000zc}
  I.~Oda,
  Phys.\ Lett.\  B {\bf 496}, 113 (2000)
  [arXiv:hep-th/0006203].

\bibitem{Ponton:2000gi}
  E.~Ponton and E.~Poppitz,
  JHEP {\bf 0102}, 042 (2001)
  [arXiv:hep-th/0012033].


\bibitem{Oda2000a} I. Oda, Phys. Rev. D {\bf 62} 126009 (2000) [hep-th/0008012].

\bibitem{Duan:2006es}
  Y.~-S.~Duan, Y.~-X.~Liu and Y.~-Q.~Wang,
  Mod.\ Phys.\ Lett.\ A {\bf 21}, 2019 (2006)
  [hep-th/0602157].



\bibitem{Gogberashvili:2007gg}
  M.~Gogberashvili, P.~Midodashvili and D.~Singleton,
  JHEP {\bf 0708}, 033 (2007)
  [arXiv:0706.0676 [hep-th]].

\bibitem{Garriga:2004tq}
  J.~Garriga and M.~Porrati,
  JHEP {\bf 0408}, 028 (2004)
  [hep-th/0406158].


\bibitem{Tinyakov:2001jt}
  P.~Tinyakov and K.~Zuleta,
  Phys.\ Rev.\ D {\bf 64}, 025022 (2001)
  [hep-th/0103062].

\bibitem{Bostock:2003cv}
  P.~Bostock, R.~Gregory, I.~Navarro and J.~Santiago,
  Phys.\ Rev.\ Lett.\  {\bf 92}, 221601 (2004)
  [hep-th/0311074].

\bibitem{Kanno:2004nr}
  S.~Kanno and J.~Soda,
  JCAP {\bf 0407}, 002 (2004)
  [hep-th/0404207].

\bibitem{Candelas:1989js}
  P.~Candelas and X.~C.~de la Ossa,
  Nucl.\ Phys.\  B {\bf 342}, 246 (1990).

\bibitem{Greene:1995hu}
  B.~R.~Greene, D.~R.~Morrison and A.~Strominger,
  Nucl.\ Phys.\  B {\bf 451}, 109 (1995)
  [arXiv:hep-th/9504145].

\bibitem{p} J.Polchinski, String theory and beyond, vol.2, Cambrigde University press.




\bibitem{Minasian:1999tt}
  R.~Minasian and D.~Tsimpis,
  Nucl.\ Phys.\ B {\bf 572}, 499 (2000)
  [hep-th/9911042].





\bibitem{Klebanov:2007us}
  I.~R.~Klebanov and A.~Murugan,
  JHEP {\bf 0703}, 042 (2007)
  [hep-th/0701064].

\bibitem{Cvetic:2000mh}
  M.~Cvetic, H.~Lu and C.~N.~Pope,
  Nucl.\ Phys.\  B {\bf 600}, 103 (2001)
  [arXiv:hep-th/0011023].

\bibitem{VazquezPoritz:2001zt}
  J.~F.~Vazquez-Poritz,
  JHEP {\bf 0209}, 001 (2002)
  [arXiv:hep-th/0111229].

\bibitem{Pando Zayas:2000sq}
  L.~A.~Pando Zayas and A.~A.~Tseytlin,
  JHEP {\bf 0011}, 028 (2000)
  [arXiv:hep-th/0010088].

\bibitem{Papadopoulos:2000gj}
  G.~Papadopoulos and A.~A.~Tseytlin,
  Class.\ Quant.\ Grav.\  {\bf 18}, 1333 (2001)
  [hep-th/0012034].

\bibitem{Klebanov:1999tb}
  I.~R.~Klebanov and E.~Witten,
  Nucl.\ Phys.\ B {\bf 556}, 89 (1999)
  [hep-th/9905104].


\bibitem{Klebanov:2000hb}
  I.~R.~Klebanov and M.~J.~Strassler,
  JHEP {\bf 0008}, 052 (2000)
  [arXiv:hep-th/0007191].

\bibitem{Klebanov:2000nc}
  I.~R.~Klebanov and A.~A.~Tseytlin,
  Nucl.\ Phys.\ B {\bf 578}, 123 (2000)
  [hep-th/0002159].




\bibitem{Firouzjahi:2005qs}
  H.~Firouzjahi and S.~H.~Tye,
  JHEP {\bf 0601}, 136 (2006)
  [arXiv:hep-th/0512076].


\bibitem{Noguchi:2005ws}
  T.~Noguchi, M.~Yamaguchi and M.~Yamashita,
  Phys.\ Lett.\  B {\bf 636}, 221 (2006)
  [arXiv:hep-th/0512249].

\bibitem{Brummer:2005sh}
  F.~Brummer, A.~Hebecker and E.~Trincherini,
  Nucl.\ Phys.\  B {\bf 738}, 283 (2006)
  [arXiv:hep-th/0510113].




\bibitem{RandjbarDaemi:2000ft}
  S.~Randjbar-Daemi and M.~E.~Shaposhnikov,
  Phys.\ Lett.\ B {\bf 491}, 329 (2000)
  [hep-th/0008087].

\bibitem{deCarlos:2003nq}
  B.~de Carlos and J.~M.~Moreno,
  JHEP {\bf 0311}, 040 (2003)
  [arXiv:hep-th/0309259].


\bibitem{Silva:2011yk}
  J.~E.~G.~Silva, C.~A.~S.~Almeida,
  Phys.\ Rev.\ D {\bf 84}, 085027 (2011)
  [arXiv:1110.1597 [hep-th]].

\bibitem{Silva:2012yj}
  J.~E.~G.~Silva, V.~Santos and C.~A.~S.~Almeida,
  Class.\ Quant.\ Grav.\  {\bf 30}, 025005 (2013)
  [arXiv:1208.2364 [hep-th]].


\end{thebibliography}
\end{document}